\RequirePackage{fix-cm}
\documentclass[smallextended]{svjour3}       
\smartqed  
\usepackage{graphicx}

\usepackage{amssymb}
\usepackage{graphicx}
\usepackage{dcolumn}
\usepackage{bm}
\usepackage{epsfig}
\usepackage{amssymb} 
\usepackage{color}
\usepackage{datetime}
\usepackage{wasysym}
\usepackage{slashed} 
\usepackage[mathscr]{euscript}
\usepackage{cancel}
\usepackage[hidelinks]{hyperref} 

\newcommand{\bal}{\begin{align}}
\newcommand{\eal}{\end{align}}
\newcommand{\beq}{\begin{eqnarray}}
\newcommand{\eeq}{\end{eqnarray}}
\newcommand{\nneeq}{\nonumber \end{eqnarray}}

\newcommand{\nn}{\nonumber \\}
\newcommand{\es}{& = &}
\newcommand{\rs}{\, = \,}

\newcommand{\cH}{ {\cal H} }

\newcommand{\cU}{ {\cal U} }
\newcommand{\cL}{ {\cal L} }

\newcommand{\h}{ {1 \over 2} }
\newcommand{\tdelta}{\tilde\delta}

\newcommand{\ket}[1]{ {\left|{#1}\right\rangle} }
\newcommand{\bra}[1]{ {\left\langle{#1}\right|} }

\newcommand{\bmat}{\left[\begin{array}}
\newcommand{\emat}{\end{array}\right]}

\begin{document}

\title{Baryon masses estimate in heavy flavor QCD
\thanks{Presented by Mar\'ia G\'omez-Rocha at the International Conference on the Structure of Baryons, November 7-11th, Universidad Pablo de Olavide, Sevilla, Spain. 
}
}
\subtitle{An effective particle approach to hadron spectra}


\author{Mar\'ia G\'omez-Rocha \and Jai More \and Kamil Serafin 
}


\institute{Mar\'ia G\'omez-Rocha \at
              Universidad de Granada and Instituto Carlos I de F\'isica  Te\'orica y Computacional \\ 
              Avda. Fuente Nueva s/n, 18071, Granada, Spain.\\
              \email{mgomezrocha@ugr.es}           
}

\date{Received: date / Accepted: date}

\maketitle

\begin{abstract}
We apply the renormalization group procedure for effective particles (RGPEP) to the QCD eigenvalue problem for only heavy quarks. We derive the effective Hamiltonian that acts on the Fock space by solving the RGPEP equation up to second order in powers of the coupling constant. The eigenstates that contain three quarks and two or more gluons are eliminated by inserting a gluon-mass term in the component with one gluon and formulate the eigenvalue problem for baryons.
We estimate masses for $bbb$ and $ccc$ states and find that the results match the estimates obtained in lattice
QCD and in quark models.

\end{abstract}

\section{Introduction}
\label{intro}

In spite of many years of research, the issue of bound states in QCD remains to be a long-standing problem to which an exact solution is still unknown. The QCD Hamiltonian which defines the Schr\"odinger equation is full of complexities. The determination of its eigenvalues, which would lead to hadron masses and the corresponding wave functions is certainly not straightforward.  

The main difficulty concerns the fact that in quantum field theory one needs to deal with an infinite number of degrees of freedom in the bound-state equation $\hat H|\psi \rangle = E|\psi \rangle$. For a baryon, which is the case discussed in this work, the eigenstate has the following structure in terms of Fock components
\begin{eqnarray}
 |\psi \rangle = |3Q\rangle + |3QG\rangle + |3QGG\rangle  + \dots \ ,
\end{eqnarray}
where we have denoted $|3Q\rangle \equiv |QQQ\rangle$; and there is no limit in the number of particles allowed. 

In this context, the renormalization group procedure for effective particles (RGPEP) was formulated as a non-perturbative tool to construct bound states in quantum field theory. 
Nonetheless, any candidate for a basic physical theory requires, an initial perturbative search for the set of interaction terms that provides the basis on which the full effective theory can be constructed in a series of successive approximations~\cite{pRGPEP}.

The RGPEP has its origin in the similarity renormalization (SRG) group for Hamiltonians~\cite{SRG1,SRG2} but, in addition, it introduces the concept of effective particles~\cite{pRGPEP,AF}. The renormalization-group approach allows one to consider particle interactions and phenomena at different energy scales. The key idea is that it is possible to express the initial Hamiltonian through a unitary similarity transformation in a scale-dependent operator basis, in such a way that for a certain scale, the number of non-negligible Fock components is small. The eigenstates depend on the renormalization-group parameter $t$ too: 
\begin{eqnarray}
 |\psi_t \rangle = |3Q_t\rangle + |3Q_tG_t\rangle + |3Q_tG_tG_t\rangle  + \dots \ .
 \label{state}
\end{eqnarray}
If an infinite number of components can be neglected in Eq.~(\ref{state}), the bound-state equation is enormously simplified and one can attempt to seek numerical solutions to the equation. 

The RGPEP has been applied and solved exactly in several simple theories~\cite{scalarmassmixng,fermionmassmixing,Glazek:2021vnw}. But, for the complex case of QCD, only perturbative expressions of the Hamiltonian have been considered so far. Second-order calculations with the inclusion of a gluon-mass ansatz have allowed us to examine the effective potential between heavy quarks in mesons~\cite{QbarQ} and in baryons~\cite{Serafin_Baryons}. Third-order calculations have been employed to calculate the running coupling in the front-form Hamiltonian~\cite{AF}, and recently, a new regularization procedure that includes a canonical gluon mass has led to analogous results~\cite{Galvez-Viruet:2022tqb}.

In this contribution, we focus on the study of triply heavy baryon spectra with equal quark masses, $ccc$ and $bbb$. For interested readers, we refer to a detailed analysis of $ccb$ and $bbc$ states provided in~\cite{Serafin_Baryons}. 

The next section presents the most important general steps in the RGPEP approach. In Section~\ref{bound} we derive the bound state equation for a system of three heavy quarks and provide the analytical result in Section~\ref{Coulomb}. The numerical setup is presented in Section~\ref{Numerical} and the corresponding results are commented in
Section~\ref{comment}. Finally,  Section~\ref{Conclusion} concludes the article.

\section{ Key elements of the RGPEP }

The starting point of this method is the Lagrangian density of the chosen theory. In this particular case, we choose QCD, $\cL_{\rm{QCD}}$. The classical Hamiltonian, $H_{\rm{QCD}}$, can be derived using Noether's theorem to calculate the energy-momentum tensor.

We use the front-form of dynamics~\cite{Dirac1949,Brodsky-Pauli-Pinsky}. In this form, four vectors are represented as $x^\mu=(x^+, x^-, x^\perp)$, where $x^+=x^0+x^3$, $x^-=x^0-x^3$, $x^\perp=(x^1,x^2)$, and the scalar product in Minkowski space-time is given by $a\cdot b=\h a^+ b^- + \h a^- b^+ - a^\perp b^\perp$. 

The quantum Hamiltonian $\hat H^{\rm{can}}_{\rm{QCD}}$ 
is derived using canonical quantization with the initial conditions on the hypersurface $x^+=0$ and in the light-cone gauge $A^+=0$. It can be expressed by the 
``-" component of the four-momentum operator, $\hat P^\mu$, which is the  generator of space-time translations~\cite{Brodsky-Pauli-Pinsky}:
\begin{eqnarray}
 \hat H_{\rm{QCD}}^{\rm{can}} = \hat P^- = \int dx^-d^2x^\perp \, : \hat \cH_{x^+=0} \,: \ .   
\end{eqnarray}
The dots on both sides of the Hamiltonian density, $\hat \cH_{x^+=0}$, indicate normal ordering of creation and annihilation operators.
In the sequel we use the notation considered in~\cite{Brodsky-Pauli-Pinsky,QQ1,QbarQ}.

The canonical Hamiltonian needs regularization and counterterms.
The regularized canonical Hamiltonian with counterterms is called \textit{initial} Hamiltonian, since it provides the initial condition for solving the RGPEP equation.

The regularization is provided by inserting functions defined in Ref.~\cite{QbarQ} in every interaction vertex. Such functions depend on ultraviolet and small-$x$ cutoffs, $\Delta$ and $\delta$, respectively, which will be removed at the end of the calculation. 

The RGPEP provides a means for the calculation of counterterms.
It introduces effective particle operators related by a unitary transformation
\begin{eqnarray}
 q_s \es \cU_s q_0 \cU_s^\dagger   \ ,
\end{eqnarray}
where $s$ has units of length and plays the role of a renormalization group parameter. It is associated with the \textit{size} of the effective particles. If $q$ and $q^\dagger$ are operators that annihilate or create pointlike particles in the Fock space, effective particle operators $q_s$ and $q_s^\dagger$ annihilate or create particles of size $s$. It is convenient to consider scale parameter $\lambda=1/s$ which has units of energy and the parameter $t=s^4$ which we have already used in Eq.~(\ref{state}).

The renormalization-group parameter labels a family of equivalent Hamiltonians that correspond to the same theory but expressed in terms of degrees of freedom that are differently defined. If $H_0=\cH_0(q_0)$ is the initial Hamiltonian, then the RGPEP demands that: 
\begin{eqnarray}
\cH_0(q_0) \es \cH_t(q_t)  \ .
\end{eqnarray}

An effective Hamiltonian that satisfies this condition is a solution of the RGPEP equation: 
\begin{eqnarray}
\cH_t' &=& [[\cH_f,\cH_{Pt}],\cH_t]  \ , 
\label{eq:rgpep}
\end{eqnarray}
where $\cH_t=\cH(q_0)$, and the prime on the effective Hamiltonian, $\cH_t$, indicates differentiation with respect to $t$. The subscript $f$ stands for \textit{free} and refers to terms that do not depend on the coupling constant. Finally,  $\cH_{Pt}$ is identical to $\cH_t$ but multiplied by a factor $\h\left(\sum_i p_i^{+}\right)^2$, with $i$ refering to all incoming and outgoing particles involved in a vertex~\cite{pRGPEP}.  

The ease of solving Eq.~(\ref{eq:rgpep}) depends on the complexity of the initial Hamiltonian. Although there are theories for which it is possible to find exact solutions~\cite{fermionmassmixing,scalarmassmixng,Glazek:2021vnw}, the complexity of QCD forces us to use a perturbative expansion in powers of the coupling constant. 
Such form of a solution can be written as
\begin{eqnarray}
\cH_t 
&=&
\cH_0 + g\cH_{t1} + g^2\cH_{t2} + g^3\cH_{t3} + g^4\cH_{t4} + \dots 
\label{Hexpansion}
\end{eqnarray}
The numerical subscript 0, 1, 2, ... refers to the power of the coupling constant. Thus, $\cH_0\equiv \cH_f$ is the 0th-order term which does not depend on $g$ the coupling constant or on the renormalization-group parameter $t$.
In the 2nd-order expansion one has
\begin{eqnarray}
\cH_{0}' + g\cH_{t1}' + g^2\cH_{t2}'  
=
\left[\left[\cH_0,\cH_0 + g\cH_{Pt\, 1} + g^2\cH_{Pt\, 1} \right],\cH_0 + g\cH_{t1} + g^2\cH_{t2} \right] 
\end{eqnarray}
which can be solved order by order:
\begin{eqnarray}\cH_0' \es 0 \ , \label{H0} \\
g\cH_{t\,1}' 
\es \left[\left[ \cH_0, g\cH_{Pt\, 1}\right],\cH_0\right] \ , \label{H1}\\
g^2\cH_{t\,2}' 
\es \left[\left[ \cH_0, g^2\cH_{Pt\, 2}\right], \cH_0\right] + \left[\left[ \cH_0, g\cH_{Pt\, 1}\right],g \cH_{1 t}\right] \ . \label{H2}
\end{eqnarray}

Solving Eqs.~(\ref{H0})-(\ref{H2}) yields exponentials of products of $t$ by differences of invariant masses. These functions play the role of form factors that appear at interaction vertices. The renormalized Hamiltonian is determined by the initial condition that at $t=0$ it should equal the regularized canonical Hamiltonian plus counterterms. The counterterms should be such that every matrix element of the renormalized Hamiltonian is cutoff independent for $t>0$, i.e. free of ultraviolent divergences.
In this work, we restrict our calculation to second-order expansions.

Note that this perturbative expansion is made at the level of the RGPEP equation, not at the level of the Schr\"odinger equation. 

\section{Effective Hamiltonian and bound-state equation for triply heavy baryons}
\label{bound}
The simplest possible systems that can be considered in QCD are heavy quarkonia and triply-heavy baryons. Thus, we simplify the picture by considering only heavy flavors, and neglecting light quarks. The eigenvalue problem simplifies enormously choosing the renormalization-group parameter in the following region
\begin{eqnarray}
 m_Q \gg \lambda \gg\Lambda_{\rm{QCD}} \ ,  
 \label{eq:hierarchyOfScales}
\end{eqnarray}
where $m_Q$ is the quark mass. The fact that $\lambda$ is much larger than $\Lambda_{\rm{QCD}}$ allows one to keep only the first term in the Hamiltonian expansion in powers of $g_t$~\cite{AF}. The condition $m_Q\gg\lambda$, on the other hand, makes Fock sectors with extra quark-antiquark pairs strongly suppressed by RGPEP form factors and they can be neglected. 
However, sectors with more gluons cannot be neglected, since they are massless, and many of them can be produced without adding much to the invariant mass of a system. 
We cannot deal with infinitely many Fock sectors of gluons. 
To address this problem, we drop all the sectors with more than one gluon and account for their absence by introducing a gluon mass ansatz in the sector $Q_t\bar Q_tG_t$ for mesons and $Q_tQ_tQ_tG_t$ for baryons~\cite{QbarQ,Serafin_Baryons}. 
Thus, our gluon mass ansatz accounts for all possible non-Abelian effects that we cannot take into account explicitly. Higher-order calculations should be able to replace such an ansatz by elements of the theory. 

The triply-heavy baryon bound-state problem with two Fock sectors (i.e. in the second order in the RGPEP) and gluon mass ansatz is (cf.~\cite{QbarQ} and~\cite{Serafin_Baryons} for more details):
\begin{eqnarray}
\left\{
\left[ \begin{array}{ll}
       (H_{t\,0} + \mu_t^2) \ \   & \ \  g H_{t1}  \\  
       g H_{t1} & \ (H_{t\,0} + g^2 H_{t2})
       \end{array} \right]
- E 
\right\}
\left[  \begin{array}{l} 
        | 3Q_t \, G_t \rangle   \\
        | 3Q_t  \rangle 
        \end{array} \right] 
\rs 0 \  .
\label{eq:matrix}
\end{eqnarray}
where $\mu_t^2$ is the gluon-mass operator, which acts on the $Q_tQ_tQ_tG_t$ sector. We assume that the mass ansatz depends on the relative motion of the gluon with respect to the quarks in that sector. 

Since we consider only terms up to second order in powers of the coupling constant in the effective Hamiltonian, the approximate eigenvalue problem Eq.~(\ref{eq:matrix}) can be reduced to the sector with no gluons~\cite{Wilson1970}. Matrix elements after the reduction are
\begin{eqnarray}
&&\bra{l} H_{\rm{eff}\,t} \ket{r} \\
&&
=
\bra{l} \left[
  H_{t0}
+ g^2 H_{t2} + \frac{1}{2} g H_{t1} 
\left(
    \frac{1}{E_l - H_{t0} - \mu_t^2}
  + \frac{1}{E_r - H_{t0} - \mu_t^2}
  \right) g H_{t1}
\right] \ket{r}
\ .
\nonumber
\label{Heff2nd}
\end{eqnarray}
where left (\textit{l}) and right (\textit{r}) states
are both in the $3Q_t$ sector and $H_f|l\rangle= E_l|l\rangle$ and $H_f|r\rangle= E_r|r\rangle$, where $H_f$ is the \textit{free} term of the Hamiltonian, which does not depend on the coupling constant. 

\section{Result: Coulomb and harmonic-oscillator potentials}
\label{Coulomb}
The effective front-form eigenvalue equation for baryons has the following structure
\begin{eqnarray}
    H_{{\rm eff}\,t}|3Q_t\rangle
\es
\frac{M^2 + P^{\perp 2}}{P^+} |3Q_t\rangle
\ ,
\end{eqnarray}
where the state $|3Q_t\rangle$ is defined as
\begin{eqnarray}
    \ket{3Q_t}
\rs
\int_{123}
P^+
\tdelta_{P.123} 
\,\psi_t(123) 
\,\frac{\epsilon^{c_1c_2c_3}}{\sqrt{6}}
b_{t\,1}^\dag b_{t\,2}^\dag b_{t\,3}^\dag
\ket{0}
\ .
\end{eqnarray}
where the spin-momentum wave function, $\psi_t(123)$ is 
multiplied by the color factor $\epsilon^{c_1c_2c_3}/\sqrt{6}$. We have used the shortcut notation $\tilde \delta_{P.123} = 2(2\pi)^3\delta^3(P-p_1+p_2-p_3)$ for the delta function of momentum conservation.
Details of the structure of the effective Hamiltonian can be found in~\cite{Serafin_Baryons}.

In the non-relativistic limit, the expressions of the interaction potentials and mass functions simplify enormously. To define this limit we introduce variables defined in~\cite{GlazekCondensatesAPP,TrawinskiConfinement},
\begin{eqnarray}
    K_{12}^\perp
    \rs
    \sqrt{\frac{x_1 + x_2}{6 x_1 x_2}}
    \kappa_{12}^\perp
    \ ,
    \quad
    K_{12}^z
    \rs
    \sqrt{\frac{x_1 + x_2}{6 x_1 x_2}}
    \frac{x_1 - x_2}{x_1 + x_2} m_Q
    \ ,
    \\
    Q_{12}^\perp
    \rs
    \sqrt{\frac{2/9}{x_3 (1 - x_3)}}
    \kappa_3^\perp
    \ ,
    \quad
    Q_{12}^z
    \rs
    \sqrt{\frac{2/9}{x_3 (1 - x_3)}}
    (3 x_3 - 1) m_Q
    \ .
\end{eqnarray}
where $\kappa_{12}^\perp$ is the relative transverse momentum of particle $1$ with respect to particle $2$, $\kappa_3^\perp$ is the relative transverse momentum of particle $3$ with respect to particles $1$ and $2$, and $x_1$, $x_2$, $x_3$ are longitudinal momentum fractions $x_i=p_i^+/P^+$ of particles $i=1$, $2$, and $3$, respectively. In the nonrelativistic approximation the eigenvalue equation can be written in the form
\begin{eqnarray}
\left[
  \frac{ {\vec K_{12}}^{\, 2} }{ 2\mu_{12} }
+ \frac{ {\vec Q_{3}}^{\, 2}  }{ 2\mu_{3(12)} }
- B
+ 3\frac{ \delta m_{1\,t}^2 }{ 2 m_Q }
\right]\,
\psi_t(123)
\nn
+ \sum_{\sigma_{1'}\sigma_{2'}}\int\frac{d^3 K_{12}'}{(2\pi)^3} \, 
  [  f_{t\,12.1'2'} V^{12}_{C,BF}
   + W^{12} ] 
\, \psi_t(1'2'3)
\nn
+ \sum_{\sigma_{2'}\sigma_{3'}}\int\frac{d^3 K_{23}'}{(2\pi)^3} \, 
  [  f_{t\,23.2'3'} V^{23}_{C,BF}
   + W^{23} ] 
\, \psi_t(12'3')
\nn
+ \sum_{\sigma_{3'}\sigma_{1'}}\int\frac{d^3 K_{31}'}{(2\pi)^3} \, 
  [  f_{t\,31.3'1'} V^{31}_{C,BF}
   + W^{31} ] 
\, \psi_t(1'23')
&= 0 \ ,    
\label{eqNR}
\end{eqnarray}
where $B$ is the binding energy;  { $V^{ij}_{C,BF} = V_{C,BF}( \vec K_{ij}, \vec K_{ij}'\, )$}
and $W^{ij} = W( \vec K_{ij}-\vec K_{ij}'\, )$ are, respectively, 
the Coulomb term with Breit-Fermi (BF) corrections and the 
additional interaction resulting from the gluon mass ansatz. 
$\mu_{12} = m_Q/2$, $\mu_{3(12)} = 2m_Q/3$ are the reduced masses.
Both $V$ and $W$ are similar to the ones in the quarkonium 
case~\cite{QbarQ}.
\begin{eqnarray}
&&V_{C,BF}(\vec K, \vec K')
=
- \frac{2}{3} g^2\,
  \frac{1}{\Delta\vec K^2}
  (1+BF)
\ ,
\\
&&W(\Delta\vec K)
=
- \frac{2}{3} g^2\left[
    \frac{1}{(\Delta K^z)^2}
  - \frac{1}{\Delta\vec K^2}
  \right]
  \frac{\mu^2}{\mu^2 + \Delta\vec K^2}
  \exp\left[-2 t m_Q^2\frac{\Delta\vec K^4}{(\Delta K^z)^2}\right]
\ \quad \quad
\label{eqforW}
\end{eqnarray}
where $\Delta\vec K = \vec K - \vec K'$ and the RGPEP form factor is
\beq
f_{t\,ij.i'j'}
\es
\exp\left\{-16 t [\vec K_{ij}^2 - (\vec K_{ij}')^2]^2\right\} \ .
\eeq
We assume now that the mass ansatz $\mu^2$ dominates $\Delta \vec K^2$ in the relevant integration range, then ${\mu^2\over \mu^2+\Delta \vec K^2} \approx 1$, and the wave function can be expanded in a Taylor series in such a way that the resulting potential that corrects the Coulomb term is a harmonic oscillator one, with oscillator frequencies
\begin{eqnarray}
\omega_{\rm{baryon}}
\es
\frac{\sqrt{3}}{2} \sqrt{\frac{\alpha}{18\sqrt{2\pi}}}
\
\frac{\lambda^3}{m_Q^2}
\ .
\label{obequal}   
\end{eqnarray}
The result of quarkonia and triply heavy baryons differ by a factor $\sqrt{3}/2$, in such a way that $\omega_{\rm{baryon}}^2 /
\omega_{\rm{meson}}^2  =3/4$.

\section{Numerical studies}
\label{Numerical}

To provide numerical results for heavy quarkonia and baryons, we need to estimate the values of $\alpha$, $m_c$, and $m_b$. Although the RGPEP is defined to provide an exact effective Hamiltonian independent of the scale parameter $\lambda$, the second-order approximation results in a certain $\lambda$-dependence, which is negligible in a window of values of $\lambda$ (cf. Figures in Refs.~\cite{Glazek:1997gt,GlazekMlynikAF}).

We assume that the scale parameter is proportional to the coupling constant and the quark mass, $\lambda\sim \sqrt{\alpha}\, m_Q$. 
In this way, if $\alpha$ is sufficiently small, the assumption satisfies our hierarchy of scales Eq.~(\ref{eq:hierarchyOfScales}) and, for the quark-antiquark system 
$\lambda \gg k_B\sim \alpha \mu$, 
where $k_B$ is the Bohr momentum and $\mu$ is the quark reduced mass. 
This ensures that the RGPEP form factors are approximately 1 and do not influence significantly the eigenvalue problem\footnote{Note that form factors are necessary in higher-order calculations since they regulate terms that otherwise would be divergent.}~\cite{Serafin_Baryons}. Furthermore, the fact that $\lambda$ is proportional to $\sqrt{\alpha}$ makes the resulting hadron binding energies proportional to $\alpha^2$, in analogy to QED~\cite{Wilsonetal}. 
In fact, the harmonic oscillator frequencies obtained from QCD are expected to be comparable in size with the strong-interaction Rydberg-like constant $R=\mu(4\alpha/3)^2/2$, since the low-mass quarkonium spectra can be characterized as intermediate between the Coulomb and the oscillator spectra~\cite{Eichten:1978tg}. 

Several simplifications are taken into account in this pilot application of our method. The numerical sketch we provide yields approximate results restricted to the low-mass hadron spectrum. 
In this numerical sketch, we estimate the Coulomb effects in first-order perturbation theory around the oscillator solution. Therefore, we consider only the diagonal matrix elements of the Coulomb potential in the basis of harmonic oscillator wave functions, since the effects of  the non-diagonal ones are relatively small and do not change significantly the lowest-mass heavy-baryon spectrum. In particular, those effects are smaller than the effects due to spin-dependent interactions, which we neglect (yet we include the effects of the Pauli exclusion principle). 

\subsection{Adjustment of parameters}

The coupling constant $\alpha$ depends on the scale parameter $\lambda$ in the following way~\cite{AF}:
\begin{eqnarray}
\alpha
\es 
\left[ \beta_0 \log \left({\lambda^2\over \Lambda_{\rm{QCD}}}\right)\right]^{-1}    
\end{eqnarray}
with $\beta_0=(33-2n_f)/(12\pi)$. We take $n_f=2$, for two heavy flavors, $b$ and $c$, though the result does not change significantly for $n_f$ in the range up to 5.
The value of $\Lambda_{\rm{QCD}}=371$~MeV is imposed by the fact that $\alpha=0.1181$ for $\lambda=M_Z=91.1876$~GeV. 

 Quark masses, $m_b$ and $m_c$, and the renormalization group parameter $\lambda$
are determined by the fit of computed heavy quarkonia spectra to the known
experimental ones. The numerical results of this fitting are given in Appendix~\ref{parameters}. Hence, our estimates of the baryon masses are predictions without any free parameters.

We would like to point out that our estimates are in a primitive stage of development. However, 
obtained results in this crude approximation are in surprisingly good agreement with other long-standing and widely used approaches. 
The purpose of this preliminary study is to find out whether the oscillator terms that follow from the assumption of gluon mass are capable of reproducing a good approximation to the heavy hadron spectrum. This would motivate higher-order studies in our perturbative expansion Eq.~(\ref{Hexpansion}), to provide a theoretical explanation of the gluon-mass generation. 

\begin{figure}
\includegraphics[width=0.9\textwidth]{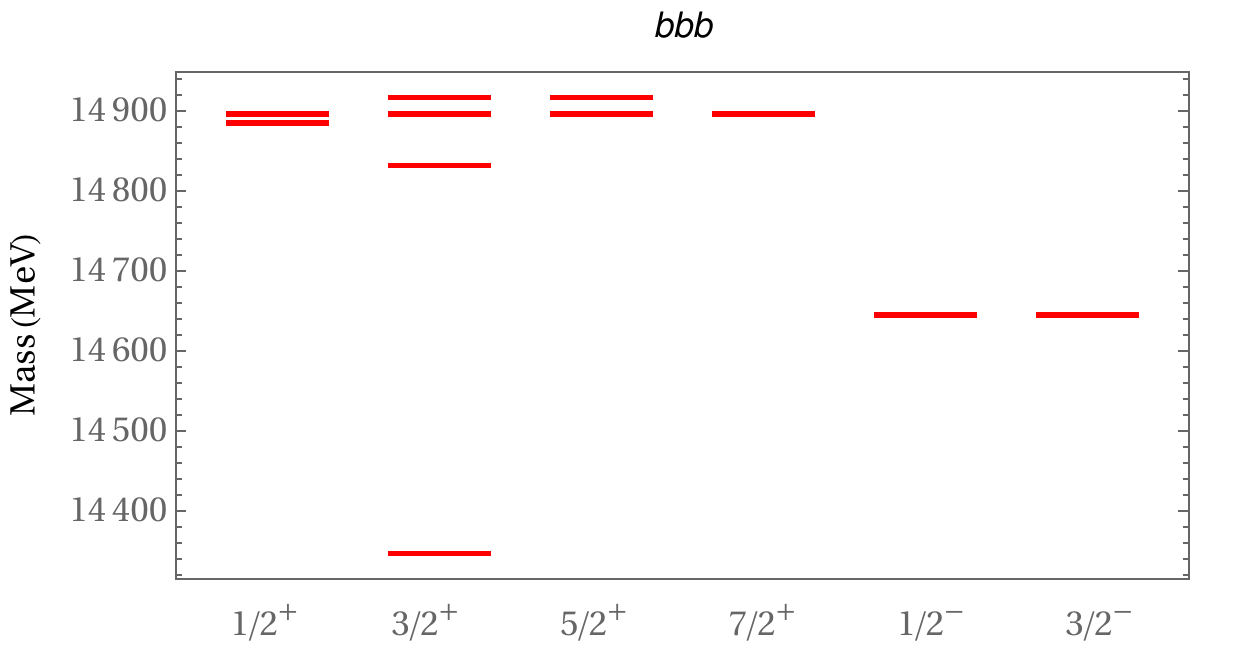}
\includegraphics[width=0.9\textwidth]{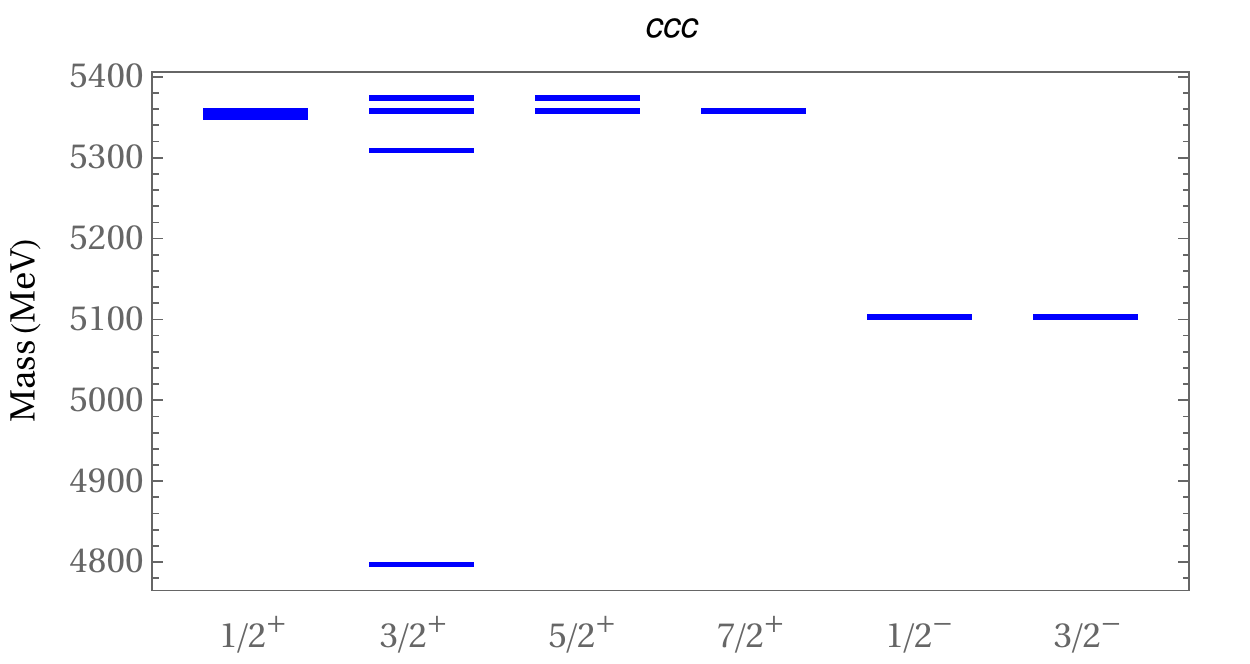}
\caption{Graphical representation of mass spectra for all $bbb$ and $ccc$ states up to second band of harmonic oscillators.}
\label{fig:bbb}       
\end{figure}

Therefore, we ignore Breit-Fermi spin interactions and estimate Coulomb effects by evaluating the expectation values of the corresponding interaction terms in the oscillator eigenstates. In this paper we select the most remarkable results and present them in Table~\ref{tab:1} and Figure~\ref{fig:bbb}.

 \begin{center}
     \begin{table}
\caption{Masses in MeV for all $bbb$ and $ccc$ states up to the second excited band of the harmonic oscillator.}
\label{tab:1}       
\begin{tabular}{ccc}
& \textbf{$bbb$-states} & \\
\hline\noalign{\smallskip}
$J^P$ & Name & Mass   \\
\noalign{\smallskip}\hline\noalign{\smallskip}
$1/2^+$ & $B_{1/2^+}$ & 14885 \\
       & $C_{1/2^+}$  &  14896 \\
\noalign{\smallskip}\hline\noalign{\smallskip}
$3/2^+$  &  $0\omega$        &  14347  \\
         &  $A_{3/2^+}$ &  14832 \\
         &  $C_{3/2^+}$ &  14896  \\
         &  $D_{3/2^+}$ &  14917 \\
\noalign{\smallskip}\hline\noalign{\smallskip}
$5/2^+$  & $C_{5/2^+}$  &  14896 \\
         & $D_{5/2^+}$  &  14917 \\
\noalign{\smallskip}\hline\noalign{\smallskip}
$7/2^+$  & $C_{7/2}$    &  14896  \\
\noalign{\smallskip}\hline\noalign{\smallskip}
 $1/2^-$  & $1\omega$  &  14645 \\
\noalign{\smallskip}\hline\noalign{\smallskip}
 $3/2^-$  & $1\omega$ &  14645 \\
\noalign{\smallskip}\hline
\end{tabular}
\hspace{1cm}
\begin{tabular}{ccc}
&\textbf{ $ccc$-states }& \\
\hline\noalign{\smallskip}
$J^P$ & Name & Mass   \\
\noalign{\smallskip}\hline\noalign{\smallskip}
$1/2^+$ & $B_{1/2^+}$ & 5350 \\
       & $C_{1/2^+}$  & 5358 \\
\noalign{\smallskip}\hline\noalign{\smallskip}
$3/2^+$  &  $0\omega$        &  4797  \\
         &  $A_{3/2^+}$ &  5309  \\
         &  $C_{3/2^+}$ &  5358  \\
         &  $D_{3/2^+}$ &  5374  \\
\noalign{\smallskip}\hline\noalign{\smallskip}
$5/2^+$  & $C_{5/2^+}$  &  5358 \\
         & $D_{5/2^+}$  &  5374 \\
\noalign{\smallskip}\hline\noalign{\smallskip}
$7/2^+$  & $C_{7/2}$    &  5358  \\
\noalign{\smallskip}\hline\noalign{\smallskip}
 $1/2^-$  & $1\omega$  &  5103 \\
\noalign{\smallskip}\hline\noalign{\smallskip}
 $3/2^-$  & $1\omega$  &  5103 \\
\noalign{\smallskip}\hline
\end{tabular}

\end{table}
\end{center}

\section{Analysis of results}
\label{comment}

The notation used is the following. 
State $0\omega$ is the ground state of the system, while state $1\omega$ is the first (orbitally) excited state.
States called $A$, $B$, $C$, and $D$ in $bbb$ and $ccc$ refer to the second excitation of the harmonic oscillator with excitation energy $2\omega$ (with $\omega\equiv\omega_ {\rm{baryon}}$) above the ground state.
They are mixtures of radial and orbital excitations and their masses differ due to different expectation values of the Coulomb potential.

The values of masses obtained for $bbb$ and $ccc$ baryons agree well with model calculations 
\cite{Bjorken:1985ei,Tsuge:1985ei,SilvestreBrac:1996bg,Jia:2006gw,Martynenko:2007je,Roberts:2007ni,Vijande:2015faa}
including quark-diquark~\cite{Giannuzzi:2009gh}
and hypercentral approximations~\cite{Ghalenovi:2014swa,Shah:2017jkr}, 
Regge phenomenology
\cite{Wei:2015gsa,Wei:2016jyk},
bag models
\cite{Ponce:1978gk,Hasenfratz:1980ka,Bernotas:2008bu},
pNRQCD
\cite{LlanesEstrada:2011kc},
sum rules~\cite{Zhang:2009re,Wang:2011ae,Aliev:2012tt,Aliev:2014lxa},
Dyson-Schwinger approach~\cite{QinQQQ}
and lattice studies~\cite{Brown:2014ena,Meinel:2012qz,Briceno:2012wt,Padmanath:2013zfa,Namekawa:2013vu,Alexandrou:2014sha,Can:2015exa}.

We remark the comparison of the ground state of $ccc$, for which different lattice approaches yield values between 4733 to 4796~MeV. Our result, 4797~MeV, differs by 29~MeV from the average result, 4768~MeV, which corresponds to the 0.6\%. In the case of the ground state of $bbb$, we obtain 14347~MeV, as compared with the lattice result of 14369~MeV, a difference of 23~MeV, 0.2\%.
Concerning mass splittings, we differ in about 10\% with lattice results provided in~\cite{Meinel:2012qz} for $bbb$ states, and in 20\% with lattice results~\cite{Padmanath:2013zfa} for the case of $ccc$ states.
Analysis of results for states $bbc$ and $ccb$ is not presented in this document. The reader is invited to consult the detailed analysis provided in~\cite{Serafin_Baryons}.

It is surprising that this preliminary approximation of our RGPEP method with no free parameters, after fitting quark masses and scale to heavy quarkonia spectra, produces similar splittings to those obtained from advanced calculations.

\section{Conclusion}
\label{Conclusion}

The effective Hamiltonian for heavy quarkonia and triply-heavy baryons derived in the second-order of our RGPEP with gluon-mass ansatz leads to baryon mass spectra that are comparable with the expectation obtained from other approaches to physics of $bbb$ and $ccc$. 

The considered method is invariant under boosts and in principle appears capable of providing a relativistic description of hadrons in terms of a small number of effective constituents, with suitably adjusted size. 
Therefore, an extension to higher-order calculations appears worth doing. 
A fourth-order calculation is needed to verify if the introduced gluon-mass ansatz provides an adequate representation of the gluon dynamics in the presence of heavy quarks. Furthermore, such calculations are also needed in the study of spin splittings and rotational symmetry. 
As a remark, it should be pointed out that the ratio $\sqrt{8/6}$ of harmonic oscillator frequencies in heavy quarkonia and baryons is close to the ratio $\sqrt{8/5}$ obtained for $u$ and $d$ quarks in constituent models using the gluon condensate. This suggests studying if the RGPEP formalism can be applied also to light hadrons as built from constituent quarks and massive gluons. 
Even in the heavy-quarks case, the effective oscillator potential provides simple wave functions that can be used  in relativistic processes involving heavy hadrons.

\begin{acknowledgements}
We thank Stanis\l aw D. G\l azek for fruitful discussions and acknowledge financial support by the FEDER2020 funds, project ref. A-FQM-406-UGR20 and from MCIN/AEI/10.13039/501100011033, Project Ref. PID2020-114767GB-I00. Figures 1 and 2 are made using the open soft-
ware JaxoDraw~\cite{Jaxodraw} distributed under the GNU General Public license. J. M. would like to thank the Department of Science and Technology (DST), Government of India, for financial
support through Grant No. SR/WOS-A/PM-6/2019(G).
\end{acknowledgements}

\appendix

\section{Masses of quarks and other parameters}
\label{parameters}

Quark masses and scale parameters are calculated using a fit to the masses of $\Upsilon(1S)$, $\Upsilon(2S)$, and $\chi_{b1}(1P)$ for bottomonia and $J/\psi$, $\psi(2S)$ and $\chi_{c1}(1P)$ for charmonia. 
Results of the fit yield:
\begin{eqnarray}
m_b\es 4698 \ \rm{MeV} \quad \rm{and} \quad \lambda_{b\bar b}=4258 \ \rm{MeV}  \ , \\
m_c\es 1460 \ \rm{MeV} \quad \rm{and} \quad \lambda_{c\bar c}=1944 \ \rm{MeV}  \ ,    
\end{eqnarray}
These values are associated, respectively, with
\begin{eqnarray}
\alpha(\lambda_{b\bar b}) = 0.2664  \quad \rm{and} \quad \omega_{b\bar b} = 268.8 \ \rm{MeV} \ , \\
\alpha(\lambda_{c\bar c}) = 0.3926  \quad \rm{and} \quad \omega_{c\bar c} = 321.6 \ \rm{MeV} \ . 
\end{eqnarray}

\bibliographystyle{unsrt}      
\bibliography{RGPEPrefs}   

\end{document}